\pdfoutput=1
\documentclass[12pt,preprint]{aastex}
\usepackage{graphicx}
\usepackage{natbib}
\usepackage{aas_macros}
\usepackage[section] {placeins}
\usepackage{subfigure}
\usepackage{float}
\usepackage{color}

\def\msun{{\rm\,M_\odot}}

\def\msun{{\rm\,M_\odot}} 

\def\zsun{{\rm\,Z_\odot}}

\newcommand{\kms}{\, {\rm km\, s}^{-1}}

\newcommand{\lya}{Ly$\alpha$ }

\def\h2{${\rm\,H_2}$}

\makeatletter 

\def\kms{{\rm\,km/s}}
\def\msun{{\rm\,M_\odot}}

\def\vol#1  {{{#1}{\rm,}\ }}
\def\lya{{\rm Ly}\alpha}

\newcount\refno
\newcount\rfno
\def\eq{$^{\the\refno\ }$\advance\refno by 1}
\def\ad{\advance\rfno by 1}

\def\clock{\count0=\time \divide\count0 by 60
     \count1=\count0 \multiply\count1 by -60 \advance\count1 by \time
     \number\count0:\ifnum\count1<10{0\number\count1}\else\number\count1\fi}

\def\myputfigure#1#2#3#4#5%
{\hskip0.03\textwidth\vskip#5pt
\makebox[0pt]{\hskip#2in
\includegraphics[width=#3\textwidth]{#1}}\vskip#4pt\hfill}

\newcount\refno
\newcount\rfno
\def\eq{$^{\the\refno\ }$\advance\refno by 1}
\def\ad{\advance\rfno by 1}

\definecolor{burntorange}{rgb}{1,0.4,0.2}

\begin{document}
\title{Composition of Low Redshift Halo Gas}

\author{Renyue Cen\altaffilmark{1}}  

\begin{abstract}

Halo gas in low-z ($z<0.5$) $\ge 0.1L_*$ galaxies in high-resolution, large-scale cosmological hydrodynamic simulations
is examined with respect to three components: (cold, warm, hot) with temperatures equal to $(<10^5$, $10^{5-6}$, $>10^6$)K,
respectively.
The warm component is compared, utilizing O~VI $\lambda \lambda$1032, 1038 absorption lines,
to observations  
and agreement is found with respect to the galaxy-O~VI line correlation, the ratio of O~VI line incidence rate in blue to red galaxies
and the amount of O~VI mass in star-forming galaxies.
A detailed account of the sources of warm halo gas 
(stellar feedback heating, gravitational shock heating and accretion from the intergalactic medium), 
inflowing and outflowing warm halo gas metallicity disparities
and their dependencies on galaxy types and environment is also presented.
Having the warm component securely anchored, our simulations make the following additional predictions.
First, cold gas is the primary component in inner regions,
with its mass comprising 50\% of all gas within galacto-centric radius $r=({\color{red}30},{\color{blue}150})$kpc
in ({\color{red} red}, {\color{blue} blue}) galaxies. 
Second, at $r>({\color{red}30},{\color{blue}200})$kpc in ({\color{red}red}, {\color{blue}blue}) galaxies
the hot component becomes the majority.
Third, the warm component is a perpetual minority, 
with its contribution peaking at {\color{blue}$\sim 30\%$} at $r=100-300$kpc in {\color{blue} blue} galaxies
and never exceeding {\color{red}5\%} in {\color{red} red} galaxies.
The significant amount of cold gas in low-z early-type galaxies found in simulations, 
in agreement with recent observations (Thom et al.) 
is intriguing,
so is the dominance of hot gas at large radii in blue galaxies.

\end{abstract}


\altaffiltext{1}{Department of Astrophysical Sciences, Princeton
  University, Peyton Hall, Ivy Lane, Princeton, NJ 08544; cen@astro.princeton.edu}

\section{Introduction}

Galaxy formation and evolution is the central astrophysical problem in cosmology.
The basic parameters of the cosmological framework -
the standard cosmological constant-dominated cold dark matter (DM) model (LCDM)
\citep[e.g.,][]{1995Krauss,1999Bahcall} - 
are largely fixed to an accuracy of $\sim 10\%$ or better.
The LCDM model is able to explain a variety of observations on scales greater than $\sim 1$Mpc,
including high redshift supernovae \citep[e.g.,][]{1998Perlmutter, 1998Riess, 2006Astier},
the cosmic microwave background \citep[e.g.,][]{2011Komatsu,2013Planck},
large-scale distribution of galaxies \citep[e.g.,][]{Tegmark04, Percival07},
X-ray cluster abundance \citep[e.g.,][]{2008Allen} and $\lya$ forest \citep[e.g.,][]{Croft02b, 2005Seljak}.

An important component of the astrophysical problem -  gravitational formation and evolution of halos that
host galaxies - is well understood, through N-body simulations \citep[e.g.,][]{2001Jenkins, 2001Bullock, 2002Wechsler, 2007Diemand}
and analytic models \citep[e.g.,][]{1991Bond, 1993LaceyCole, Sheth99, Mo02, Cooray02}.
The gastrophysics of galaxy formation and feedback, on the other hand, is far from being adequately understood.
Alternative approaches that parameterize and then infer physical processes based on finding
best matches to observations, such as  the semi-analytic methods \citep[e.g.,][]{Som99b, 2003Benson}
and the halo-occupation distribution (HOD) method \citep[e.g.,][]{2002Berlind, Zheng07},
have been successful but have limited predictive power.
More importantly, in semi-analytic methods the treatment of galaxy formation is halo based and 
largely decoupled from that of the intergalactic medium, which in fact has dramatically evolved with time.
At $z=2-6$ most of the baryons are found to be in the 
$\lya$ forest, a relatively cold phase of temperature of $\sim 10^4$K, 
as indicated by both observations \citep[e.g.,][]{1997Rauch} and simulations \citep[e.g.,][]{1994Cen}. 
By $z=0$ most of the baryons in the intergalactic medium have been heated up, primarily by gravitational shocks, 
to temperatures that are broadly peaked at about $10^6$K, the so-called Warm-Hot Intergalactic Medium (WHIM) \citep[e.g.,][]{1999Cen}.
The ``ab initio", more predictive approach of direct cosmological hydrodynamic simulations,
after having made steady progress \citep[e.g.,][]{1994Evrard, 1996Katz, 2002Teyssier, 2005Keres,
2006Hopkins, 2006Oppenheimer, 2007Governato, 2007Naab, 2008Gnedin, 2009Joung, 2011bCen},
begin to be able to make statistically significant and physically realistic characterizations 
of the simultaneous evolution of galaxies and the intergalactic medium.

It is the aim of this writing to quantify the composition of the halo gas in low redshift galaxies,
using state-of-the-art high resolution ($460h^{-1}$pc), large-scale (thousands of galaxies) cosmological hydrodynamic simulations
with advanced treatments of star formation, feedback and microphysics.
Our focus here is on gas that is in the immediate vicinities of galaxies,
on galactocentric distances of $10-500$kpc,
where the exchanges of gas, metals, energy and momentum between galaxies
and the intergalactic medium (IGM) primarily take place.
We shall broadly term it ``circumgalactic medium (CGM)" or ``halo gas".
Understanding halo gas is necessary before a satisfactory theory of galaxy formation and evolution may be constructed.
The present theoretical study is also strongly motivated observationally, 
in light of recent rapid accumulation of data by HST observations 
enabling detailed comparisons between galaxies and the warm component ($T\sim 10^5-10^6$K) of their CGM at low redshift
\citep[e.g.,][]{2009Chen, 2011bProchaska, 2011bTumlinson, 2011Tripp}.

We shall dissect halo gas at low redshift ($z<0.5$) into three components, (cold, warm, hot) gas with temperature ($<10^5$, $10^5-10^6$, $>10^6$)K,
respectively. 
A large portion of our presentation is spent on quantifying 
O~VI $\lambda \lambda$1032, 1038 absorption lines 
and the overall properties of warm halo gas and comparing them to observations in as much detail as possible.
Feedback processes, while being treated with increased physical sophistication, are still not based on first principles
due primarily to resolution limitations in large-scale cosmological simulations.
Thus, it is imperative that our simulations are well validated and anchored by 
requiring that some key and pertinent aspects of our simulations match relevant observations.
The O~VI line, when collisionally ionized, has its abundance peaked at a temperature of $T=10^{5.3-5.7}$K 
and thus is an excellent proxy for the the warm gas.
After validating our simulations with respect to the observed properties of O~VI absorption lines,
we present the overall composition of low redshift halo gas.
We find that, for ({\color{red} red},{\color{blue}blue}) galaxies more luminous than $0.1L_*$ the cold gas of $T<10^5$K, on average, 
dominates the halo gas budget within a radius of $({\color{red}30},{\color{blue}150})$kpc. 
Beyond a radius of $({\color{red}30},{\color{blue}200})$kpc 
for ({\color{red} red},{\color{blue}blue}) galaxies 
the hot gas of $T>10^6$K dominates.
The warm component remains a smallest minority at all radii, peaking at $\sim 30$\% at $\sim 100-300$kpc for {\color{blue} blue} galaxies
but never exceeding 5\% for {\color{red} red} galaxies.

The following physical picture emerges for the physical nature of the warm gas component.
The warm halo gas has a cooling time much shorter than the Hubble time
and hence is ``transient", with their presence requiring sources.
To within a factor of two we find that, for low-z 
$\ge 0.1L_*$ 
{\color{red} red} 
galaxies
contributions to warm halo gas from
star formation feedback ($F_{\color{red} r}$), accretion of intergalactic medium ($A_{\color{red} r}$)
and gravitational shock heating ($G_{\color{red} r}$) are $(F_{\color{red} r},A_{\color{red} r},G_{\color{red} r})=(30\%, 30\%, 40\%)$.
For {\color{blue} blue} $\ge 0.1L_*$ galaxies contributions to warm halo gas from
the three sources are $(F_{\color{blue} b},A_{\color{blue} b},G_{\color{blue} b})=(48\%, 48\%, 4\%)$.
The mean metallicity of warm halo gas in ({\color{red} red}, {\color{blue} blue}) galaxies 
is ($\sim 0.25\zsun$, $\sim 0.11\zsun$).
Environmental dependence of O~VI-bearing halo gas is as follows.
In low density environments the metallicity of inflowing warm gas 
is substantially lower than that of outflowing warm gas;
the opposite is true in high density environments.

The outline of this paper is as follows.
In \S 2.1 we detail simulation parameters and hydrodynamics code,
followed by a description of our method of making synthetic O~VI spectra in \S 2.2, 
which is followed by a description of how we average the two separate simulations C (cluster) and V (void) run in \S 2.3.
Results are presented in \S 3.
A detailed comparison of galaxy-O~VI absorber correlation is computed and shown to match observations in \S 3.1,
followed in \S 3.2 by an analysis of the ratio of O~VI absorber incidence rates around {\color{blue} blue} and {\color{red} red} galaxies
that is found to be consistent with observations.
A detailed examination of the physical origin and properties of the warm gas in low-z halo is given in \S 3.3.
The overall composition of low-z halo gas is given in \S 3.4 and conclusions are summarized in \S 4.

\section{Simulations}\label{sec: sims}

\subsection{Hydrocode and Simulation Parameters}

We perform cosmological simulations with the AMR Eulerian hydro code, Enzo 
\citep[][]{1999aBryan, 1999bBryan, 2005OShea}.
The version we use is a ``branch" version \citep[][]{2009Joung},
which includes a multi-tiered refinement method that allows for spatially varying 
maximum refinement levels, when desired.
This Enzo version also includes metallicity-dependent radiative cooling extended down to $10~$K,
molecular formation on dust grains,
photoelectric heating and other features that are different from or not in the public version of Enzo code.
We use the following cosmological parameters that are consistent with 
the WMAP7-normalized \citep[][]{2011Komatsu} LCDM model:
$\Omega_M=0.28$, $\Omega_b=0.046$, $\Omega_{\Lambda}=0.72$, $\sigma_8=0.82$,
$H_0=100 h {\rm km s}^{-1} {\rm Mpc}^{-1} = 70 {\rm km} s^{-1} {\rm Mpc}^{-1}$ and $n=0.96$.
These parameters are also consistent with the latest Planck results
\citep[][]{2013Planck},
if one adopts the Hubble constant that is the average between Planck value and those
derived based on SNe Ia and HST key program \citep[][]{2011Riess, 2012Freedman}.
We use the power spectrum transfer functions for cold dark matter particles and 
baryons using fitting formulae from \citet[][]{1998Eisenstein}.
We use the Enzo inits program to generate initial conditions.

First we ran a low resolution simulation with a periodic box of $120~h^{-1}$Mpc on a side.
We identified two regions separately, one centered on
a cluster of mass of $\sim 2\times 10^{14}\msun$
and the other centered on a void region at $z=0$.
We then resimulate each of the two regions separately with high resolution, but embedded
in the outer $120h^{-1}$Mpc box to properly take into account large-scale tidal field
and appropriate boundary conditions at the surface of the refined region.
We name the simulation centered on the cluster ``C" run
and the one centered on the void  ``V" run.
The refined region for ``C" run has a size of $21\times 24\times 20h^{-3}$Mpc$^3$
and that for ``V" run is $31\times 31\times 35h^{-3}$Mpc$^3$.
At their respective volumes, they represent $1.8\sigma$ and $-1.0\sigma$ fluctuations.
The root grid has a size of $128^3$ with $128^3$ dark matter particles.
The initial static grids in the two refined boxes
correspond to a $1024^3$ grid on the outer box.
The initial number of dark matter particles in the two refined boxes
correspond to $1024^3$ particles on the outer box.
This translates to initial condition in the refined region having a mean interparticle-separation of 
$117h^{-1}$kpc comoving and dark matter particle mass of $1.07\times 10^8h^{-1}\msun$.
The refined region is surrounded by two layers (each of $\sim 1h^{-1}$Mpc) 
of buffer zones with 
particle masses successively larger by a factor of $8$ for each layer, 
which then connects with
the outer root grid that has a dark matter particle mass $8^3$ times that in the refined region.
The initial density fluctuations 
are included up to the Nyquist frequency in the refined region.
The surrounding volume outside the refined region 
is also followed hydrodynamically, which is important in order to properly capture
matter and energy exchanges at the boundaries of the refined region.
Because we still can not run a very large volume simulation with adequate resolution and physics,
we choose these two runs of moderate volumes to represent two opposite environments that possibly bracket the universal average.

We choose a varying mesh refinement criterion scheme such that the resolution is always better than $460$/h proper parsecs 
within the refined region, corresponding to a maximum mesh refinement level of $9$ above $z=3$, 
of $10$ at $z=1-3$ and $11$ at $z=0-1$.
The simulations include a metagalactic UV background
\citep[][]{2012Haardt},  
and a model for shielding of UV radiation by atoms \citep[][]{2005Cen}.
The simulations also include metallicity-dependent radiative cooling and heating \citep[][]{1995Cen}. 
We clarify that our group has included metal cooling and metal heating (due to photoionization of metals) 
in all our studies since \citet[][]{1995Cen} for the avoidance of doubt \citep[e.g.,][]{2009Wiersma, 2011TepperGarcia}.
Star particles are created in cells that satisfy a set of criteria for 
star formation proposed by \citet[][]{1992CenOstriker}.
Each star particle is tagged with its initial mass, creation time, and metallicity; 
star particles typically have masses of $\sim$$10^{5-6}\msun$.

Supernova feedback from star formation is modeled following \citet[][]{2005Cen}.
Feedback energy and ejected metal-enriched mass are distributed into 
27 local gas cells centered at the star particle in question, 
weighted by the specific volume of each cell (i.e., weighting is equal to the inverse of density), 
which is to mimic the physical process of supernova
blastwave propagation that tends to channel energy, momentum and mass into the least dense regions
(with the least resistance and cooling).
We allow the whole feedback processes to be hydrodynamically coupled to surroundings
and subject to relevant physical processes, such as cooling and heating, as in nature.
The extremely inhomogeneous metal enrichment process
demands that both metals and energy (and momentum) are correctly modeled so that they
are transported into right directions in a physically sound (albeit still approximate 
at the current resolution) way, at least in a statistical sense.
In our simulations metals are followed hydrodynamically by solving 
the metal density continuity equation with sources (from star formation feedback) and sinks (due to subsequent star formation).
Thus, metal mixing and diffusion through advection, turbulence and other hydrodynamic processes
are properly treated in our simulations.

The primary advantages of this supernova energy based feedback mechanism are three-fold.
First, nature does drive winds in this way and energy input is realistic.
Second, it has only one free parameter $e_{SN}$, namely, the fraction of the rest mass energy of stars formed
that is deposited as thermal energy on the cell scale at the location of supernovae.
Third, the processes are treated physically, obeying their respective conservation laws (where they apply),
allowing transport of metals, mass, energy and momentum to be treated self-consistently 
and taking into account relevant heating/cooling processes at all times.
We use $e_{SN}=1\times 10^{-5}$ in these simulations.
The total amount of explosion kinetic energy from Type II supernovae
with a Chabrier IMF translates to $e_{SN}=6.6\times 10^{-6}$.
Observations of local starburst galaxies indicate
that nearly all of the star formation produced kinetic energy (due to Type II supernovae)
is used to power galactic superwinds \citep[e.g.,][]{2001Heckman}.
Given the uncertainties on the evolution of IMF with redshift (i.e., possibly more top heavy at higher redshift)
and the fact that newly discovered prompt Type I supernovae contribute a comparable
amount of energy compared to Type II supernovae, it seems that our adopted value for
$e_{SN}$ is consistent with observations and physically realistic.
The validity of this thermal energy-based feedback approach comes empirically.
In \citet[][]{2012Cen} the metal distribution in and around galaxies over a wide range of redshift
($z=0-5$) is shown to be in excellent agreement with respect to the properties of observed damped $\lya$ systems
\citep[][]{2012Rafelski},
whereas in \citet[][]{2012bCen} we further show that 
the properties of O~VI absorption lines at low redshift, including their abundance, Doppler-column density distribution,
temperature range, metallicity and coincidence between O~VII and O~VI lines,
are all in good agreement with observations \citep[][]{2008Danforth,2008Tripp, 2009Yao}.
This is non-trivial by any means, because they require that the transport of metals and energy from galaxies to
star formation sites to megaparsec scale be correctly modeled as a function of distance over the entire cosmic timeline,
at least in a statistical sense.

\subsection{Simulated Galaxy Catalogs}

We identify galaxies in our high resolution simulations using the HOP algorithm
\citep[][]{1999Eisenstein}, operated on the stellar particles, which is tested to be robust
and insensitive to specific choices of concerned parameters within reasonable ranges.
Satellites within a galaxy are clearly identified separately.
The luminosity of each stellar particle at each of the Sloan Digital Sky Survey (SDSS) five bands
is computed using the GISSEL (Galaxy Isochrone Synthesis Spectral Evolution Library) stellar synthesis code \citep[][]{Bruzual03},
by supplying the formation time, metallicity and stellar mass.
Collecting luminosity and other quantities of member stellar particles, gas cells and dark matter
particles yields
the following physical parameters for each galaxy:
position, velocity, total mass, stellar mass, gas mass,
mean formation time,
mean stellar metallicity, mean gas metallicity,
star formation rate,
luminosities in five SDSS bands (ugriz) and others.
At a spatial resolution of proper $460$pc/h with more than 2000 well resolved galaxies at $z=0$,
this simulated galaxy catalog presents an excellent (by far, the best available)
tool to study circumgalactic medium around galaxies at low reshift.

In some of the analysis we perform here 
we divide our simulated galaxy sample into two sets according to the galaxy color.
We shall call galaxies with 
{\color{blue} $g-r<0.6$ blue} and 
those with 
{\color{red} $g-r>0.6$ red}. 
It is found that $g-r=0.6$ is at the trough of the galaxy bimodal color distribution
of our simulated galaxies \citep[][]{2011bCen, 2012Tonnesen}, which 
agrees well with that of observed low-z galaxies \citep[e.g.,][]{2003Blanton}.

\subsection{Generation of Synthetic O~VI Absorbers}

The photoionization code CLOUDY (Ferland et al. 1998) is used 
post-simulation to compute the abundance of O~VI, adopting the shape of the UV
background calculated by \citet[][]{2012Haardt} normalized by the intensity at 1 Ryd determined by 
\citet[][]{1999Shull} and assuming ionization equilibrium.
We generate synthetic absorption spectra 
given the density, temperature, metallicity and velocity fields in simulations.
Each absorption line is identified by the velocity (or wavelength) interval between one downward-crossing and the next upward-crossing points
at flux equal to $0.99$ (flux equal to unity corresponds to an unabsorbed continuum flux) in the spectra.
We do not add instrumental and other noises to the synthetic spectra.
Since the absorption lines in question are sparsely distributed in velocity space,
their identifications have no significant ambiguity.
Column density, equivalent width, Doppler width,
mean column density weighted velocity and physical space locations, 
mean column density weighted temperature, density and metallicity are computed for each line.
We sample the C and V run, respectively, with $72,000$ and $168,000$ 
random lines of sight at $z=0$, 
with a total pathlength of $\Delta z\sim 2000$.
A total of $\sim 30,000$ $\ge 50$~mA O~VI absorbers are identified in the two volumes.
While a detailed Voigt profile fitting of the flux spectrum would have enabled 
closer comparisons with observations, 
simulations suggest that such an exercise does not necessarily provide a more clarifying 
physical understanding of the absorber properties, because bulk velocities are very important 
and velocity substructures within an absorber
do not necessarily correspond to separate physical entities \citep[][]{2012bCen}.

\subsection{Averaging C and V Runs}

The C and V runs at $z=0$ are used to obtain an ``average" of the universe.
This cannot be done precisely without much larger simulation volumes, which is presently not feasible.
Nevertheless, we make the following attempt to obtain an approximate average.
The number density of galaxies with luminosity greater than
$0.1L_*$ in SDSS r-band in the two runs is found to be $3.95\times 10^{-2}$$h^3$Mpc$^{-3}$ and 
$1.52\times 10^{-2}$$h^3$Mpc$^{-3}$, respectively, in the C and V box. 
We fix the weighting for C and V run for the purpose of averaging statistics of the C and V runs
by requiring that the average density of galaxies with luminosity greater than
$0.1L_*$ in SDSS r-band in the simulations to be equal to the observed global value of 
$2.87\times 10^{-2}$$h^3$Mpc$^{-3}$ by SDSS \citep[][]{2003Blanton}.
In the results shown below we use this method to obtain averages of statistics,
where doing so allows for some more quantitative comparisons with observed data.

\section{Results}

\subsection{Galaxy-O~VI Absorber Correlation at $z=0-0.5$}

\begin{figure}[h!]    
\begin{center}
\vskip -0.0cm
\centering
\hskip -0.2in
\resizebox{5.0in}{!}{\includegraphics[angle=0]{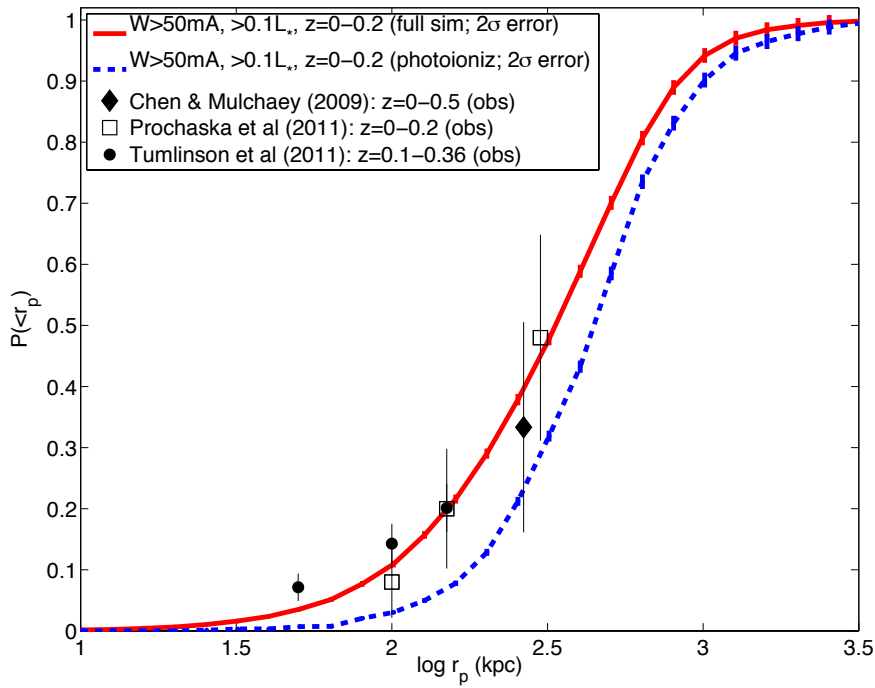}}    
\end{center}
\vskip -0.5in
\caption{
Cumulative probability distribution functions of $\ge 50$~mA O~VI absorbers
of finding $\ge 0.1L_*$ galaxies at $z=0-0.2$ from simulations with $2\sigma$ errorbars (red solid curves). 
The distribution functions at $z=0-0.2$ are obtained by averaging
$z=0$ and $z=0.2$ results with equal weighting.
Also shown as symbols are observations from 
\citet[][]{2009Chen} (solid diamonds),
\citet[][]{2011Prochaska} (open squares)
and
\citet[][]{2011Tumlinson} (solid dots).
Because the impact parameter of \citet[][]{2011Tumlinson} samples reaches only $150$kpc,
we have normalized their data points by matching their $r_p=150$kpc point to 
the $r_p=150$kpc point of \citet[][]{2011Prochaska}.
The blue dashed curve is produced when only photoionized O~VI lines with temperature $T\le 3\times 10^4$K
in our simulations are used.
The $\chi$ square per degree of freedom for the red solid curve using all observed 
data points is $1.2$, whereas it is $7.6$ for the blue dashed curve.
}
\label{fig:profLr}
\end{figure}

Figure~\ref{fig:profLr} shows 
the cumulative probability distribution functions of $\ge 50$~mA O~VI absorbers
of finding $\ge 0.1L_*$ galaxies at $z=0-0.2$ from simulations as well as observations.
We find good agreement between simulations and observations, quantified by
the $\chi$ square per degree of freedom of $1.2$.
In comparison, if using only the low temperature ($T<3\times 10^4$K) O~VI absorbers in the simulations,
the cumulative probability is no longer in reasonable agreement with observations,
with the $\chi$ square per degree of freedom equal to $7.6$;
this exercise, however, only serves as an illustration of what a photoionization dominated model may produce.
It will be very interesting to make a similar calculation directly using 
SPH simulations that have predicted the dominance of photoionized O~VI absorbers 
even for strong O~VI absorbers as shown here \citep[e.g.,][]{2012Oppenheimer}.

This significant difference found with respect to the strong O~VI absorber-galaxy cross correlations 
between the photoionization and collisional ionization dominated models
stems from the relative difference in the locations of strong O~VI absorbers in the two models.
In the collisional ionization dominated model \citep[][]{2012bCen, 2011Shull} 
the strong O~VI absorbers are spatially closer to galaxies in order to have high
enough temperature (hence high O~VI abundance) and high enough density to make strong O~VI absorbers,
whereas in the photoionization dominated model \citep[][]{2011TepperGarcia, 2012Oppenheimer}
they have to be sufficiently far from galaxies to have low enough densities
to be photoionized to O~VI.
Additional requirement in the latter for production of strong O~VI absorbers
is high metallicity ($\ge 0.1\zsun$) to yield high enough O~VI columns,
as found in SPH simulations \citep[][]{2011TepperGarcia, 2012Oppenheimer}.

\subsection{O~VI Absorbers Around Blue and Red Galaxies}

\begin{figure}[h!]    
\begin{center}
\vskip -0.0cm
\centering
\hskip -0.2in
\resizebox{6.0in}{!}{\includegraphics[angle=0]{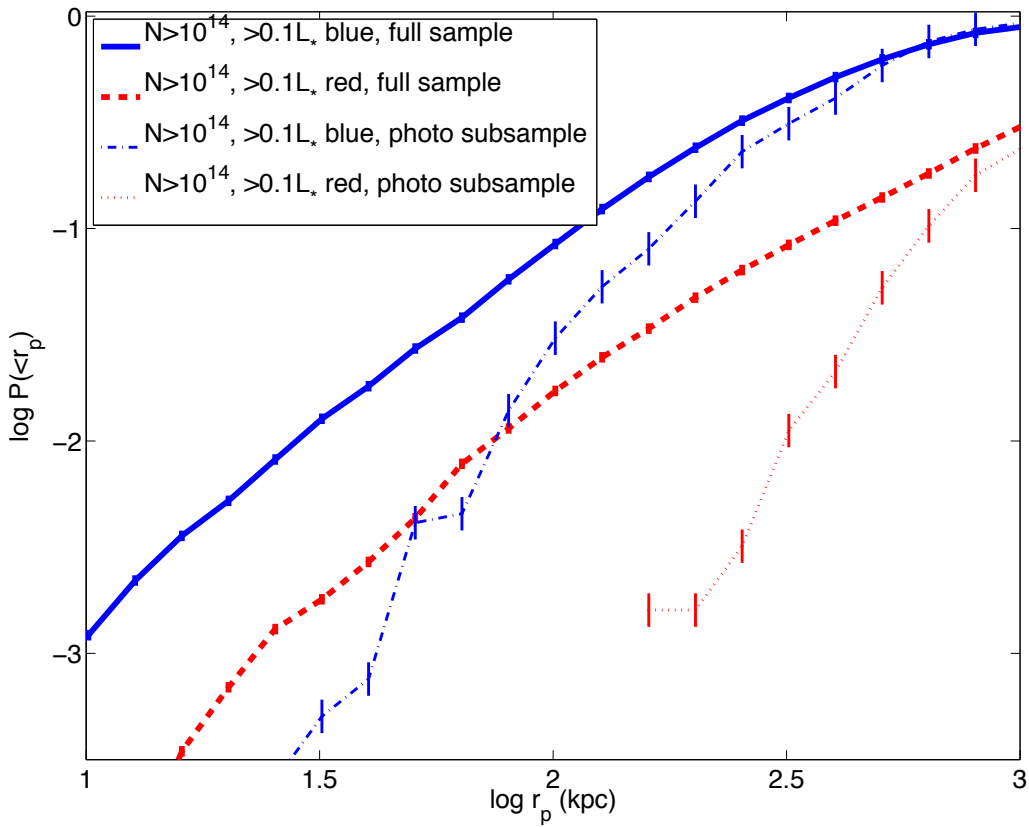}}   
\end{center}
\vskip -0.5in
\caption{
The cumulative probability distribution functions of O~VI absorbers
with column density greater than $10^{14}$~cm$^{-2}$ of finding a ({\color{red} red}, {\color{blue} blue}) galaxy of luminosity 
of $\ge 0.1L_*$ (in SDSS r-band) at $z=0.2$ with ({\color{red} red dashed}, {\color{blue} blue solid}) curves from simulations 
with $10\sigma$ errorbars. 
The ({\color{red} red dotted curve}, {\color{blue} blue dot-dashed curve}) are the corresponding functions
for the subset of O~VI absorbers that have temperature $T\le 3\times 10^4$K in our simulations.
}
\label{fig:probLrsub}
\end{figure}

\begin{figure}[ht]
\begin{center}
\vskip -0.0cm
\centering
\hskip -0.2in
\resizebox{6.0in}{!}{\includegraphics[angle=0]{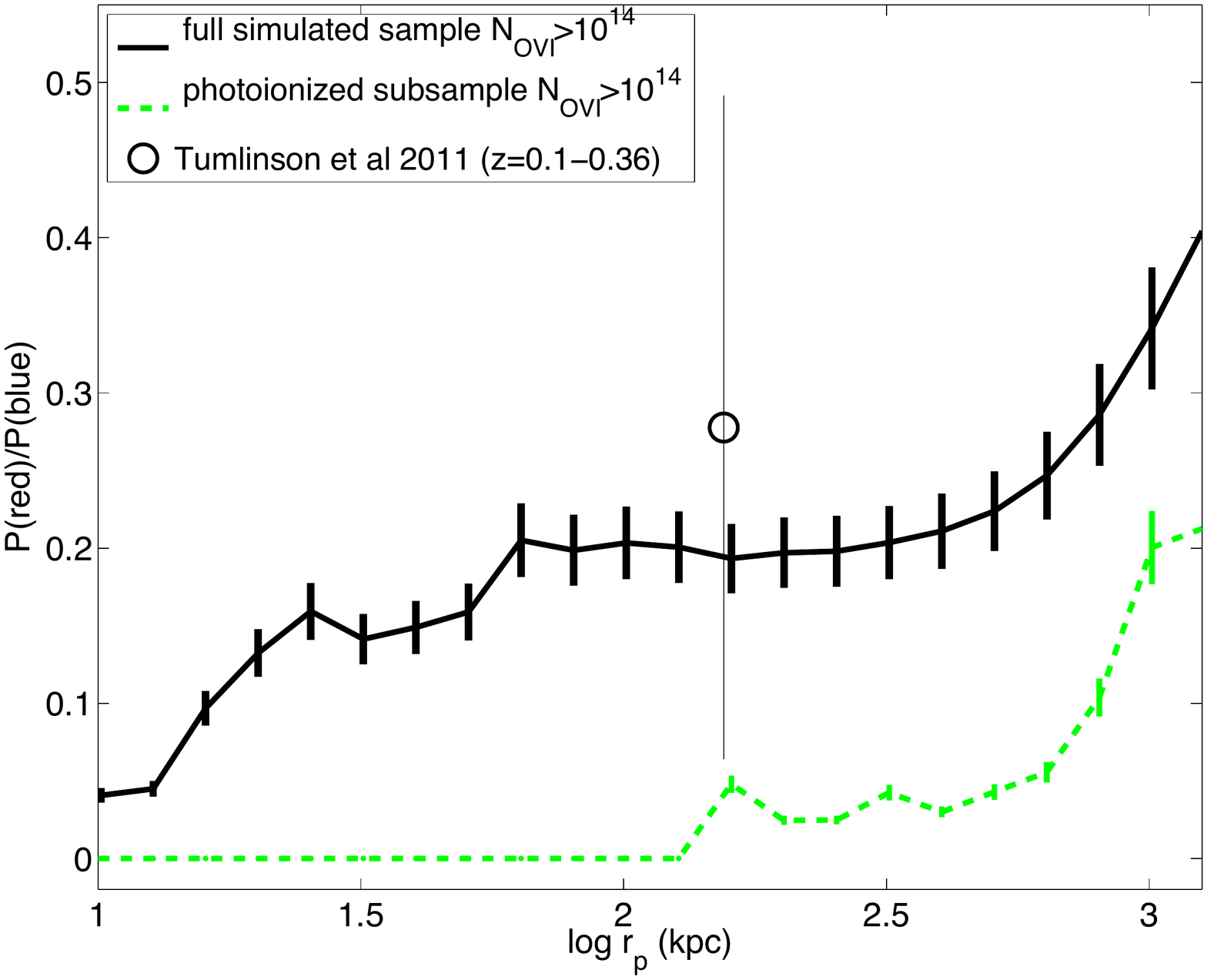}}     
\end{center}
\vskip -0.5in
\caption{
The ratio of cumulative radial probability distribution function of O~VI absorbers
of equivalent width (${\rm W}$) greater than $50$mA per {\color{red} red} galaxy to that per {\color{blue} blue} galaxy  
of $\ge 0.1L_*$ (in SDSS r-band) at redshift $z=0.2$ from simulations (the black solid curve, $2\sigma$ errorbars). 
The same ratio for photoionized O~VI absorbers ($T\le 3\times 10^4$K) only 
is shown as the green dashed curve.
Also shown as an open circle is the observation by \citet[][]{2011Tumlinson}.
}
\label{fig:ratioLrsub}
\end{figure}

Observations have shown an interesting dichotomy of O~VI incidence rate around {\color{blue} blue} and {\color{red} red} galaxies.
Figure~\ref{fig:probLrsub} shows 
the cumulative probability distribution functions of $N_{\rm OVI}>10^{14}$~cm$^{-2}$ O~VI absorbers of finding a ({\color{red} red}, {\color{blue} blue}) galaxy of luminosity 
of $\ge 0.1L_*$ at $z=0.2$ from simulations.
In Figure~\ref{fig:ratioLrsub} we show the ratio of 
the cumulative radial distribution per {\color{red} red} galaxy to 
per {\color{blue} blue} galaxy of $\ge 0.1L_*$ at $z=0.2$, compared to observations. 
It is seen in Figure~\ref{fig:ratioLrsub}
that in the $r=50-300$kpc range 
the ratio of incidence rate of strong O~VI absorbers 
around {\color{red} red} galaxies to that around {\color{blue} blue} galaxies is about 1:5, 
in quantitative agreement with observations.
In the case with photoionized O~VI absorbers only, the fraction of O~VI absorbers around {\color{red} red} galaxies
is much lower and lies significantly below the observational estimates,
although the present small observational sample prevents from reaching strong statistical conclusions 
based on this ratio alone.

We see in Figure~\ref{fig:probLrsub} that statistical uncertainties 
of the radial probability distribution of simulated O~VI absorbers are already very small 
due to a significant number of simulated galaxies and a still larger number of simulated absorbers used.
What limits the ability to make firm statistical statements
is the sample size of observational data.
Hypothetically, if the mean remains the same, a factor of two smaller errorbars 
would render the photoionization dominated model inconsistent with observations at $\ge 2\sigma$ confidence level,
whereas our collisionally dominated model would be consistent with observations within $1\sigma$.

\begin{figure}[h!]
\begin{center}
\vskip -0.0cm
\centering
\resizebox{7.25in}{!}{\includegraphics[angle=0]{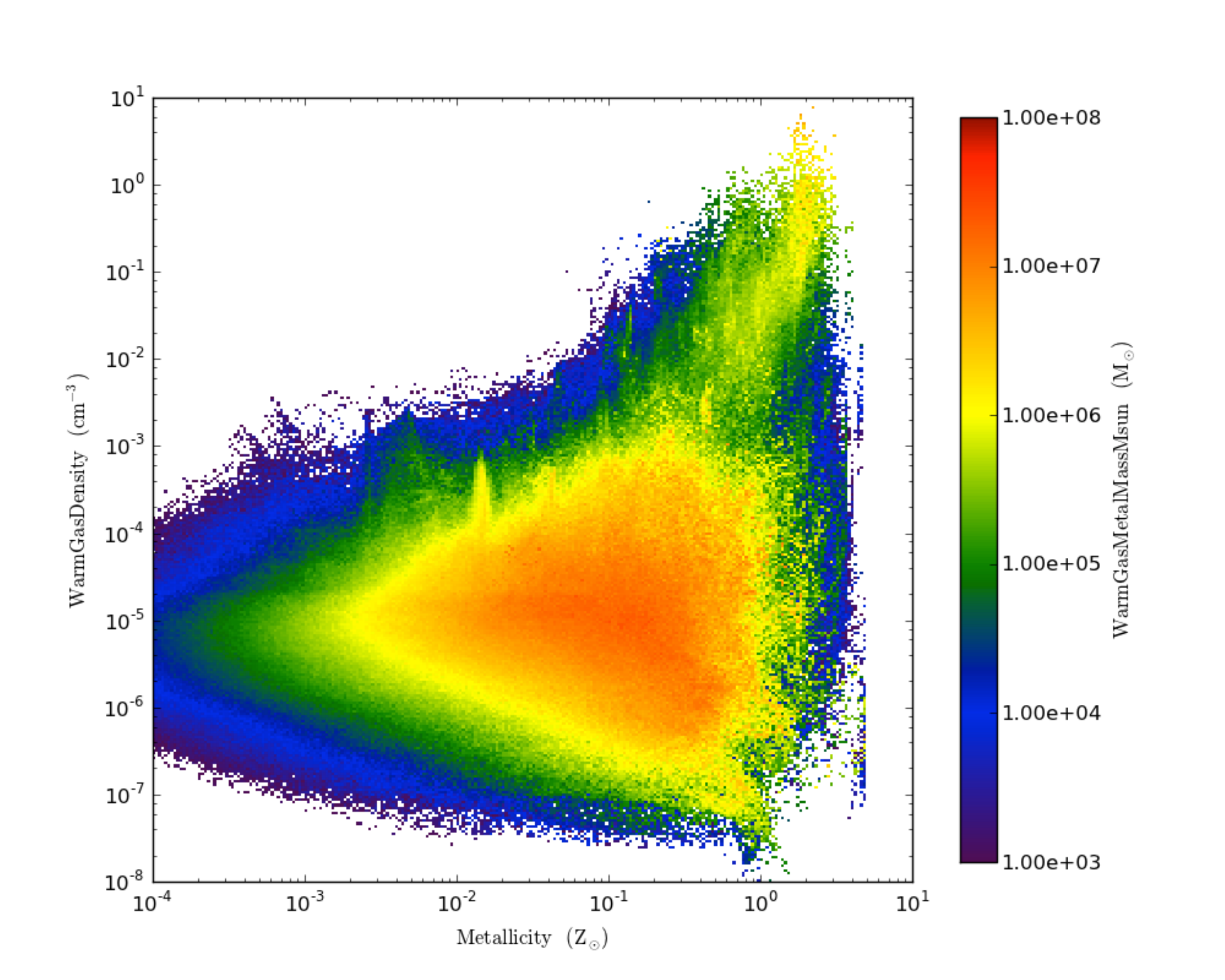}}   
\end{center}
\vskip -0.5in
\caption{
shows the metal mass in the warm gas ($T=10^{5}-10^6$K) 
distributed in the density-metallicity phase space. This is a good proxy for O~VI bearing gas.
The amount of gas mass in the WHIM is 40\% of total gas in the simulation volume.
}
\label{fig:WarmGasMetalMassonZnWarm}
\end{figure}

\subsection{Physical Origin of O~VI Absorbers}

We now turn to an analysis to give a physical 
description for the origin of O~VI absorbers in the CGM, in the context of the cold dark matter based model.
While this section is interesting on its own for physically understanding halo gas, 
the next section on halo gas mass decomposition is not predicated on it.

First, we ask whether the warm gas traced by O~VI absorbers requires significant energy input to be sustained
over the Hubble time.
Figure~\ref{fig:WarmGasMetalMassonZnWarm}
shows the metal mass in the warm gas ($T=10^{5}-10^6$K) 
distributed in the density-metallicity phase space. 
We see that most of warm metals is concentrated 
in a small phase space region centered at 
$({\rm n, Z})=(10^{-5}{\rm cm}^{-3},0.15 \zsun)$.
We note that the amount of gas mass in the WHIM is 40\% of total gas averaged over the simulation volumes,
in agreement with previous simulations \citep[e.g.,][]{1999Cen, 2001Dave, 2006Cen}
and other recent simulations \citep[][]{2011Smith,2010Dave, 2010Shen, 2010Tornatore}. 
For this gas we find that the cooling time is $t_{\rm cool}=6\times 10^8$~yrs (assuming a temperature of $10^{5.5}$K), 
shorter than the Hubble time by a factor $\ge 20$. 
For the strong O~VI absorbers considered here, the cooling time is still shorter.
In other words, either (1) energy is supplied to sustain existing O~VI gas or (2) new warm gas is accreted
or (3) some hotter gas needs to continuously cool through the warm phase.
Since O~VI gas by itself does not define a set of stable systems and 
is spatially well mixed or in close proximity with other phases of gas,
this suggests that the O~VI gas in halos is {\it ``transient"} in nature.

\begin{figure}[h!]
\begin{center}
\vskip -0.0cm
\centering
\hskip -0.2in
\resizebox{6.0in}{!}{\includegraphics[angle=0]{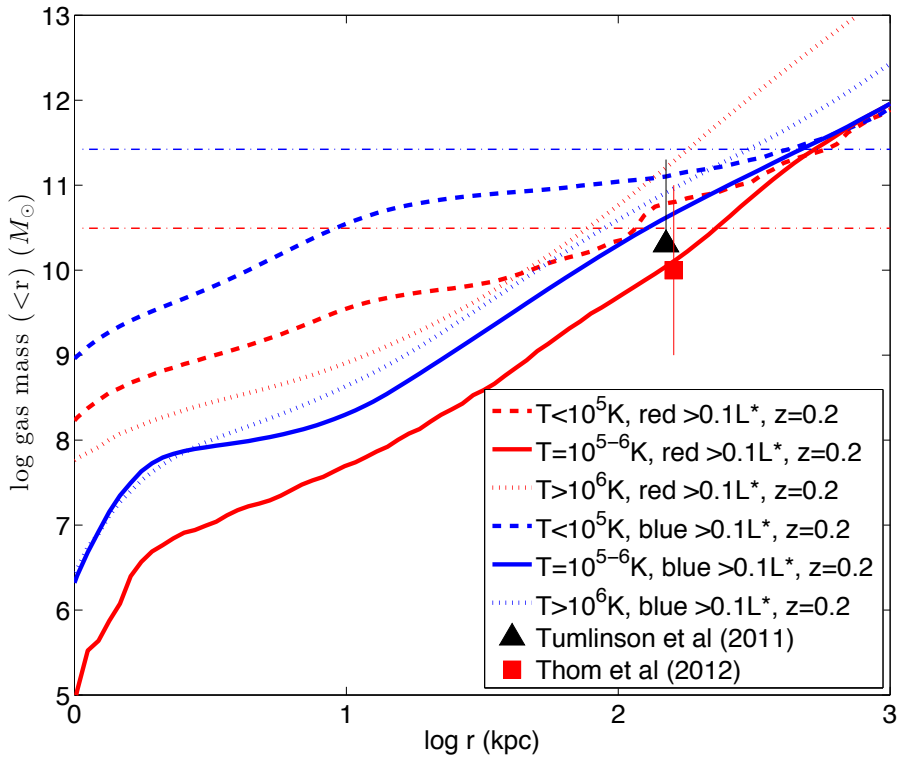}}   
\end{center}
\vskip -0.5in
\caption{
shows the cumulative gas mass as a function of radius 
for cold (dashed curves), warm-hot (solid curves)
and hot gas (dotted curves) around {\color{blue} blue (blue curves)}
and {\color{red} red (red curves)} galaxies at $z=0.2$.
Also shown the the black triangle is the 
lower limit from observations of \citet[][]{2011bTumlinson} for star forming galaxies.
The horizontal {\color{blue} blue} and {\color{red} red dot-dashed} lines are
the amount of warm gas the respective star formation rate can possibly produce.
Additional data point for cold ($T<10^5$K) gas in early-type galaxies within $150$kpc 
is also plotted as the the red square with the errorbars indicating an estimated vertical range 
from observations of \citet[][]{2012Thom} based on $15$ early-type galaxies.
}
\label{fig:Lr011e4gascompgasmass}
\end{figure}

We consider three sources of warm halo gas:
mechanical feedback energy from stellar evolution, 
gravitational binding energy released from halo formation and interactions,
and direct accretion from the IGM.
This simplification sets a framework to make a quantitative assessment 
of these three sources for warm gas that we now describe.
We denote $F_{\color{blue} b}$ and $F_{\color{red} r}$ as the O~VI incidence rate (in some convenient units) per {\color{blue} blue} and {\color{red} red} $\ge 0.1L_*$ galaxy 
due to star formation feedback energy heating,
$G_{\color{blue} b}$ and $G_{\color{red} r}$ as those due to gravitational heating,
and $A_{\color{blue} b}$ and $A_{\color{red} r}$ as those due to accreted gas from the IGM.
It is useful to stress the distinction between $G$ and $A$.
$A$ is gas directly accreted from IGM that is either already warm or heated up to be warm
by compression upon accretion onto the halo.
On the other hand, $G$ is gas that is shock heated to the warm phase or to a hotter phase that cools back down
to become warm.
Restricting our analysis to within a galactocentric radius of $150$kpc
and reading off numbers from the red curve in Figure~\ref{fig:ratioLrsub},
we obtain two relations: 
\begin{eqnarray}
\label{eq:source}
F_{\color{blue} b} + G_{\color{blue} b} + A_{\color{blue} b} = 5,  \nonumber \\
F_{\color{red} r} + G_{\color{red} r} + A_{\color{red} r} = 1. 
\end{eqnarray}
An additional reasonable assumption is now made: 
feedback heating rate $S_{\color{red} r}$ ($S_{\color{blue} b}$) is 
proportional to average star formation rate ${\rm SFR_{{\color{red} r}}}$ (${\rm SFR_{{\color{blue} b}}}$), 
which in turn is proportional to their respective gas accretion rate $A_{\color{red} r}$ ($A_{\color{blue} b}$).
This assumption allows us to lump 
$F_{\color{red} r}$ and $A_{\color{red} r}$ ($F_{\color{blue} b}$ and $A_{\color{blue} b}$):
\begin{eqnarray}
\label{eq:STsource}
S_{\color{blue} b} = F_{\color{blue} b} + A_{\color{blue} b} = C\times {\rm SFR_{\color{blue} b}}  \nonumber \\
S_{\color{red} r} = F_{\color{red} r} + A_{\color{red} r} = C\times {\rm SFR_{\color{red} r}},
\end{eqnarray}
where $C$ is a constant.
We will return to determine $F_{\color{red} r}$ and $A_{\color{red} r}$ ($F_{\color{blue} b}$ and $A_{\color{blue} b}$) separately later.
Equation (\ref{eq:source}) is now simplified to:
\begin{eqnarray}
\label{eq:source2}
S_{\color{blue} b} + G_{\color{blue} b} = 5,  \nonumber \\
S_{\color{red} r} + G_{\color{red} r} = 1. 
\end{eqnarray}
The ratio of ${\rm SFR_{\color{blue} b}}$ to ${\rm SFR_{\color{red} r}}$ can be computed directly in the simulations, found to be $8.4$. 
Rounding it down to $8$ and combining it with Equation (\ref{eq:STsource}) give
\begin{eqnarray}
\label{eq:sfr}
S_{\color{blue} b}/S_{\color{red} r}=8.
\end{eqnarray}
Lastly, a direct assessment of the relative strength of gravitational heating 
of warm gas in {\color{blue} blue} and {\color{red} red} galaxies is obtained by making the following ansatz:
the amount of hot $T\ge 10^6$K gas is proportional to the overall heating rate,
to which  the gravitational heating rate of warm gas is proportional.
Figure~\ref{fig:Lr011e4gascompgasmass} shows 
the gas mass of the three halo gas components interior to the radius shown in the x-axis.
Within the galactocentric radius of $150$kpc it is found that 
the amount of hot halo gas per {\color{red} red} $\ge 0.1L_*$ galaxy is twice that of per {\color{blue} blue} $\ge 0.1L_*$ galaxy:
\begin{eqnarray}
\label{eq:grav}
G_{\color{red} r}/G_{\color{blue} b}=2.
\end{eqnarray}
Solving Equations (\ref{eq:source2},\ref{eq:sfr},\ref{eq:grav}) yields
\begin{eqnarray}
\label{eq:solution}
S_{\color{blue} b}=24/5, G_{\color{blue} b} = 1/5;  \nonumber \\
S_{\color{red} r}=3/5, G_{\color{red} r}=2/5.
\end{eqnarray}
The estimate given in Equations (\ref{eq:grav}) is admittedly uncertain.
Therefore, an estimate on how sensitively conclusions depend on it is instructive.  
We find that, 
if we had used $G_{\color{red} r}/G_{\color{blue} b}=1$ (instead of $2$),
we would have obtained $S_{\color{blue} b}=32/7, G_{\color{blue} b} = 3/7, S_{\color{red} r}=4/7, G_{\color{red} r}=3/7$;
had we used $G_{\color{red} r}/G_{\color{blue} b}=1/2$,
we would have obtained $S_{\color{blue} b}=4, G_{\color{blue} b} = 1, S_{\color{red} r}=1/2, G_{\color{red} r}=1/2$.
Thus, a relatively robust conclusion for the sources of warm halo gas emerges:
(1) for {\color{red} red} $\ge 0.1L_*$ galaxies 
($F_{\color{red} r} + A_{\color{red} r}$) and 
$G_{\color{red} r}$ have the same magnitude, 
(2) for {\color{blue} blue} $\ge 0.1L_*$ galaxies 
($F_{\color{blue} b} + A_{\color{blue} b}$) overwhelmingly dominates over 
$G_{\color{blue} b}$.

\begin{figure}[h!]
\begin{center}
\vskip -0.0cm
\centering
\hskip -0.2in
\resizebox{5.0in}{!}{\includegraphics[angle=0]{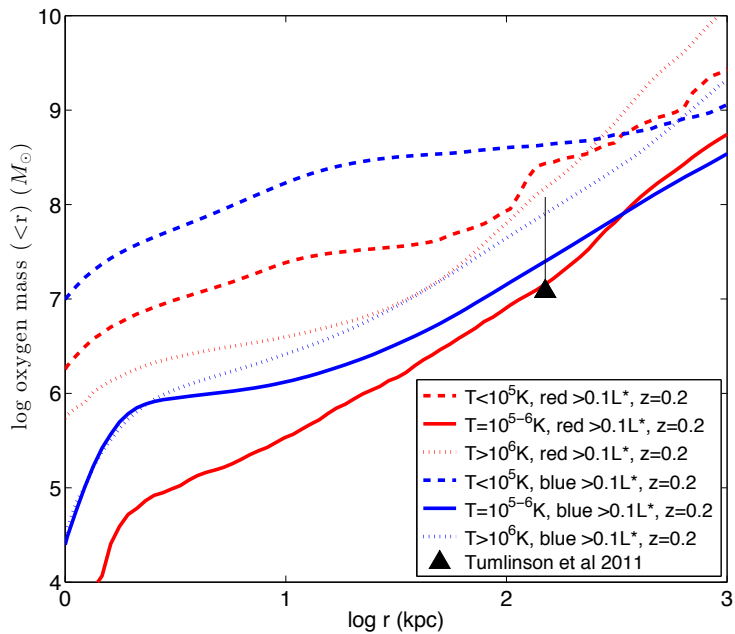}}   
\end{center}
\vskip -0.5in
\caption{
shows the cumulative metal mass as a function of radius 
for cold (dashed curves), warm-hot (solid curves)
and hot gas (dotted curves) around {\color{blue} blue (blue curves)}
and {\color{red} red (red curves)} galaxies at $z=0.2$.
}
\label{fig:Lr011e4mtlcompgasmass}
\end{figure}

It is prudent to have a consistency check for 
the conclusion that star formation feedback may dominate heating of warm gas
that produces the observed O~VI absorbers in {\color{blue} blue} galaxies.
In Figure~\ref{fig:Lr011e4gascompgasmass} the horizontal {\color{blue} blue} dot-dashed line
is obtained by assuming a Chabrier-like IMF that is used in the simulations, 
which translates to $2/3 \times 10^{-5}{\rm SFR}\times t_{\rm cool}\times c^2/(k 10^{5.5}K)$,
where $c$ is speed of light and $k$ Boltzmann constant; also assumed is that $2/3$ of the initial supernova energy 
is converted to gas thermal energy, which is the asymptotic value for Sedov explosions,
${\rm SFR}$ the respective average star formation rate per $\ge 0.1L_*$ {\color{blue} blue} galaxy,
$t_{\rm cool}=6\times 10^8$yrs an estimated cooling time for warm halo gas.
From this illustration we see that with about 20\% efficiency of heating warm gas,
star formation feedback energy is already adequate for accounting for all the observed warm gas around {\color{blue} blue} galaxies.
We therefore conclude that the required energy from star formation feedback to heat up 
the warm gas is available and our conclusions are self-consistent,
even if the direct accretion contribution is zero, which we will show is not.
Our results on warm gas mass are also in reasonable agreement with observations of \citet[][]{2011bTumlinson},
so is the oxygen mass contained in the warm component,
as shown in Figure~\ref{fig:Lr011e4mtlcompgasmass}. 

\begin{deluxetable}{llll}
\tablecolumns{4}
\tablewidth{0pc}
\tablecaption{Warm Inflow and Outflow at $r=[50-150]$kpc Radial Shell}
\tablehead{
\colhead{\ }
&\colhead{$|v_r|>0\kms$}
&\colhead{$|v_r|>100\kms$}
&\colhead{$|v_r|>250\kms$}
}
\startdata
\ & $(f_{\rm in},{\rm Z}_{\rm in}/\zsun, {\rm Z}_{\rm out}/\zsun)$ & $(f_{\rm in},{\rm Z}_{\rm in}/\zsun, {\rm Z}_{\rm out}/\zsun)$ & $(f_{\rm in},{\rm Z}_{\rm in}/\zsun, {\rm Z}_{\rm out}/\zsun)$ \\
C {\color{red} red} & (58\%, 0.27, 0.17) & (58\%, 0.29, 0.17) & (59\%, 0.31, 0.17) \\
V {\color{red} red} & (51\%, 0.21, 0.29) & (61\%, 0.18, 0.26) & (65\%, 0.11, 0.33) \\
C {\color{blue} blue} & (54\%, 0.099, 0.10) & (55\%, 0.099, 0.10) & (55\%, 0.099, 0.10) \\
V {\color{blue} blue} & (52\%, 0.10, 0.14) & (52\%, 0.09, 0.16) & (46\%, 0.08, 0.24) \\
\enddata
\tablecaption{
The first gives a letter label for each run.
The second, third and fourth columns give
the comoving box size, comoving spatial resolution
and dark matter particle mass.
The last column indicates the GSW strength.
\label{table1}}
\end{deluxetable}

Let us now determine $F_{\color{red} r}$ and $A_{\color{red} r}$ ($F_{\color{blue} b}$ and $A_{\color{blue} b}$) individually in the following way.
We compute warm metal mass that have inward and outward radial velocities within a radial shell
at $r=[50,150]$kpc separately for all {\color{red} red} $>0.1L_*$ galaxies and all {\color{blue} blue} $>0.1L_*$ galaxies, 
denoting inflow warm metal mass as $M_Z(v_r<0)$ and outflow warm metal mass as $M_Z(v_r>0)$,
where $v_r$ is radial velocity of a gas element with positive being outflowing and negative being inflowing.
We define the inflow warm metal fraction as $f_{\rm in}\equiv M_Z(v_r<0)/(M_Z(v_r<0) + M_Z(v_r>0))$,
which is listed as the first of the three elements in each entry in Table 1 under the column $|v_r|>0\kms$.
We also compute the mean metallicities (in solar units) for the inflow and outflow warm gas,
which are the second and third of the three elements in each entry in Table 1.
Four separated cases are given: 
(1) {\color{red} red} galaxies in C run (C {\color{red} red}),
(2) {\color{red} red} galaxies in V run (V {\color{red} red}),
(3) {\color{blue} blue} galaxies in C run (C {\color{blue} blue}),
(4) {\color{blue} blue} galaxies in V run (V {\color{blue} blue}).
In order to make sure that inflow and outflow are not confused with random motions of gas in a Maxwellian like distribution,
we separately limit the magnitude of infall and outflow radial velocities to greater than 
$100\kms$ and $250\kms$, and listed the computed 
quantities under the third column $|v_r|>100\kms$ 
and the fourth column $|v_r|>250\kms$, respectively.

It is interesting to first take a closer look at the difference in metallicities between inflow and outflow gas.
The warm inflow gas in {\color{red} red} galaxies in the C run has consistently higher metallicity than warm outflow gas,
$Z_{\rm in}=(0.27-0.31)\zsun$ versus $Z_{\rm out}=0.17\zsun$.
The opposite holds for {\color{red} red} galaxies in the V run: $Z_{\rm in}=(0.11-0.21)\zsun$ versus $Z_{\rm out}=(0.26-0.33)\zsun$.
The warm inflow gas in {\color{blue} blue} galaxies in the C run has about the same metallicity as warm outflow gas
at $Z=(0.09-0.1)\zsun$.
The warm inflow gas in {\color{blue} blue} galaxies in the V run, on the other hand, has a substantially 
lower metallicity than the warm outflow gas,
$Z_{\rm in}=(0.08-0.10)\zsun$ versus $Z_{\rm out}=(0.14-0.24)\zsun$.
Except in the case of C {\color{blue} blue},
we note that the inflow and outflow gas has different metallicities,
with the difference being larger when a higher flow velocitiy is imposed in the selection.
This difference in metallicity demonstrates that the {\it warm inflows and outflows are
distinct dynamical entities}, not random motions in a well-mixed gas,
making our distinction of inflows and outflows physically meaningful.
A physical explanation for the metallicity trends found can be made as follows.
In low density environment (i.e., in the V run) circumgalactic medium has not been enriched to 
a high level and hot gas is not prevalent.
As a result, warm (and possibly cold) inflows of relatively low metallicities still exist at low redshift.
The progression from {\color{blue} blue} to {\color{red} red} galaxies in the V run reflects a progression from very low
density regions (i.e., true voids) to dense filaments and group environments,  
with higher metallicities for both inflows and outflows in the denser environments in the V run; 
but the difference between inflow  and outflow metallicities remains. 
For {\color{red} red} galaxies in high density environments (C run) the circumgalactic medium has been enriched to higher metallicities. 
Higher cooling rates of higher-metallicity gas in relatively hot environments 
preferentially produces higher-metallicity warm gas that originates from hot gas and has now cooled to become warm gas. 
The {\color{blue} blue} galaxies in the C run are primarily in cosmic filaments and the metallicity of the inflow gas
is about $0.1\zsun$, which happens to coincide with the metallicity of the outflow gas.
One needs to realize that at the radial shell $r=[50-150]$kpc over which the tabulated quantities are computed,
the outflow gas originated in star forming regions 
has loaded a substantial amount of interstellar and circumgalactic medium in the propagation process.

Let us now turn to the warm inflow and outflow metal mass.
It appears that the fraction of inflow warm metals (out of all warm metals)
lies in a relatively narrow range $f_{\rm in}=45-65\%$.
For our present purpose we will just say $F_{\color{red} r}=A_{\color{red} r}$ and $F_{\color{blue} b}=A_{\color{blue} b}$.
Armed with these two relations
our best estimates for various contributions to the observed warm halo metals, as a good proxy for the O~VI absorption,
can be summarized as follows.

\noindent
$\bullet$ 
For {\color{red} red} $\ge 0.1L_*$ galaxies at $z=0.2$ contributions to warm metals in the halo gas from
star formation feedback ($F_{\color{red} r}$), accretion of intergalactic medium ($A_{\color{red} r}$)
and gravitational shock heating ($G_{\color{red} r}$) are $(F_{\color{red} r},A_{\color{red} r},G_{\color{red} r})=({\color{red} 30\%, 30\%, 40\%})$.

\noindent
$\bullet$ 
For {\color{blue} blue} $\ge 0.1L_*$ galaxies at $z=0.2$ contributions to warm metals in the halo gas from
the three sources are $(F_{\color{blue} b},A_{\color{blue} b},G_{\color{blue} b})=({\color{blue}48\%, 48\%, 4\%})$.

\noindent
$\bullet$ 
Dependencies of warm halo gas metallicities on galaxy type and environment are complex but physically understandable.
For red galaxies, the metallicity of inflowing warm gas increases with increasing environmental overdensity,
whereas that of outflowing warm gas decreases with increasing environmental overdensity.
For blue galaxies, 
the metallicity of inflowing warm gas depends very weakly on environmental overdensity,
whereas that of outflowing warm gas decreases with increasing environmental overdensity.
As a whole, the mean metallicity of warm halo gas in 
{\color{red} red} 
galaxies is 
{\color{red} $\sim 0.25\zsun$},
while that of {\color{blue} blue} 
galaxies is {\color{blue} $\sim 0.11\zsun$}.

\begin{figure}[h!]
\begin{center}
\vskip -0.0cm
\centering
\hskip -0.2in
\resizebox{4.5in}{!}{\includegraphics[angle=0]{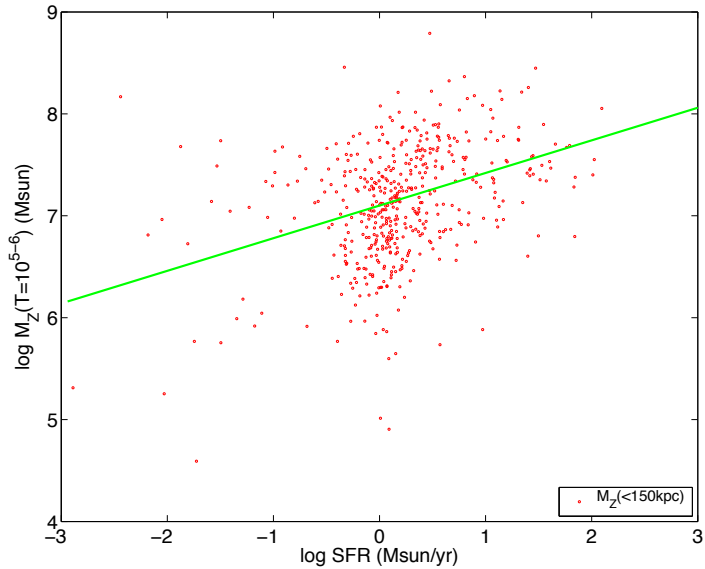}}   
\end{center}
\vskip -0.5in
\caption{
shows the metal mass in the warm gas $M_Z$ within a galactocentric radius of $150$kpc as a function of the SFR of the galaxy at $z=0.2$.
Each red dot is a galaxy.
The green curve shows the best linear regression, $\log {\rm M_Z(T=10^{5-6}K)/\msun=0.32}\log{\rm SFR} + 7.1$, for the galaxies shown.
}
\label{fig:M56SFR}
\end{figure}

\noindent
We suggest that these estimates of source fractions are not seriously in error on average, if one is satisfied with an accuracy of a factor of two.
The relative metallicity estimates should be quite robust with errors much smaller than a factor of two.
It is stressed that these estimates are averaged over many {\color{red} red} and {\color{blue} blue} galaxies 
and one is not expected to have been led to think that the correlations (such as between warm gas mass and SFR) hold strictly for individual galaxies.
Rather, we expect large variations from galaxy to galaxy, even at a fixed star formation rate.
Figure~\ref{fig:M56SFR} makes this important point clear,
which shows that, while there is a positive correlation between warm metal mass 
within $150$kpc radius and SFR for galaxies with non-negligible SFR (i.e., appearing in the SFR range shown), a dispersion of $\sim 1$ dex in warm metal mass
at a fixed SFR in the range of $0.1-100\msun$~yr$^{-1}$ exists.
The goodness of the fit can be used as a way to rephrase this significant dispersion.
If one assumes that the errorbar size is each log mass determination for each shown galaxy is 1,
one finds that the chi-square per degree of the fitting line (green) is $0.80$,
indicating that the correlation between 
$\log {\rm M_Z(T=10^{5-6}K)}$  and $\log{\rm SFR}$ is only good to about $1$ dex in warm metal gas mass.

\subsection{Composition of Low-z Halo Gas}

In \citet[][]{2012bCen} we show that 
the properties of O~VI absorption lines at low redshift, including their abundance, Doppler-column density distribution,
temperature range, metallicity and coincidence between O~VII and O~VI lines,
are all in good agreement with observations \citep[][]{2008Danforth,2008Tripp, 2009Yao}.
In the above we have shown that O~VI-galaxies relations as well as oxygen mass in galaxies in the simulations
are also in excellent agreement with observations.
These tests together are non-trivial and lend us significant confidence to now examine the overall composition of halo gas at low-z.

\begin{figure}[h!]
\begin{center}
\vskip -0.0cm
\centering
\hskip -0.45in
\resizebox{3.7in}{!}{\includegraphics[angle=0]{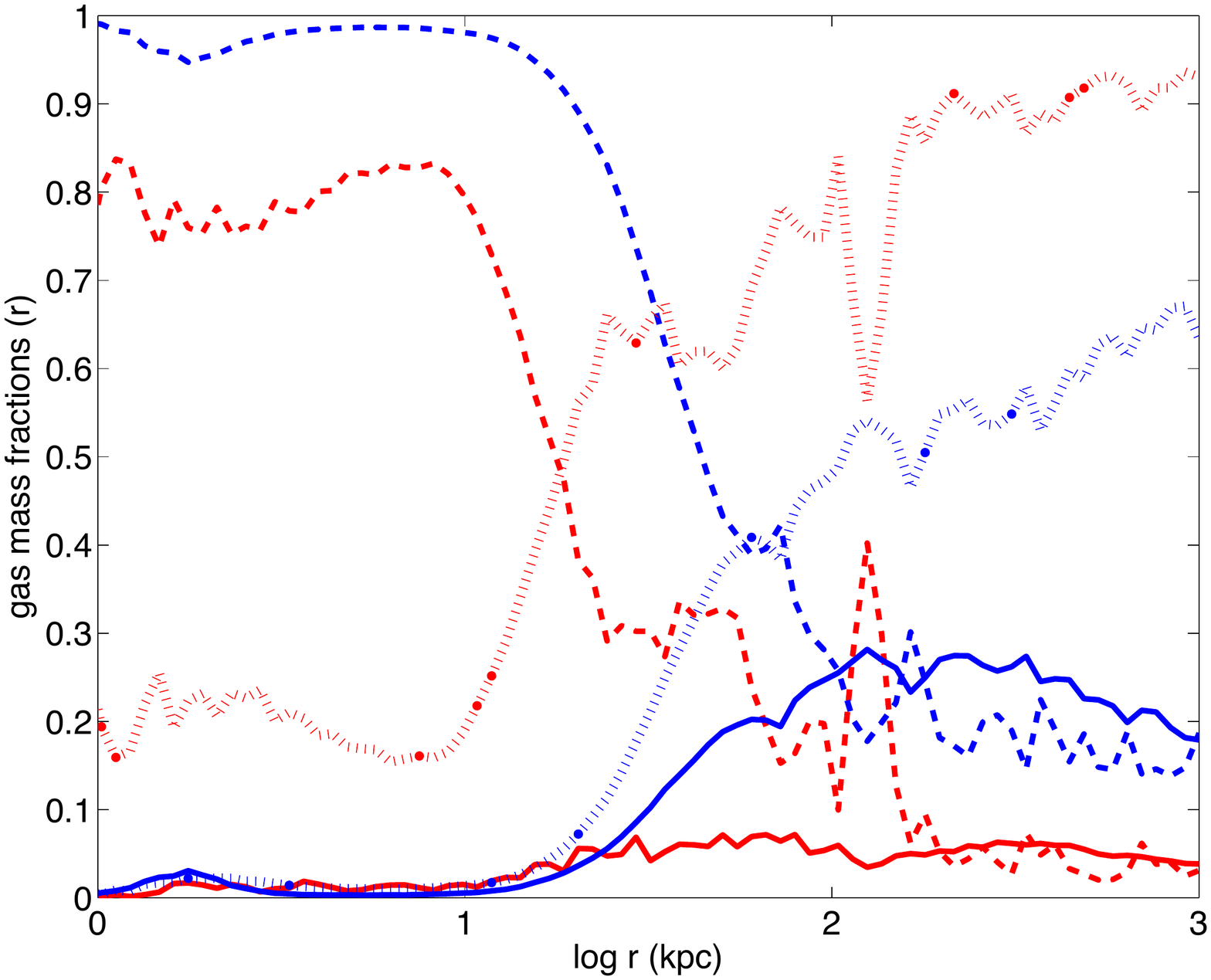}}  
\hskip -0.5in
\resizebox{3.7in}{!}{\includegraphics[angle=0]{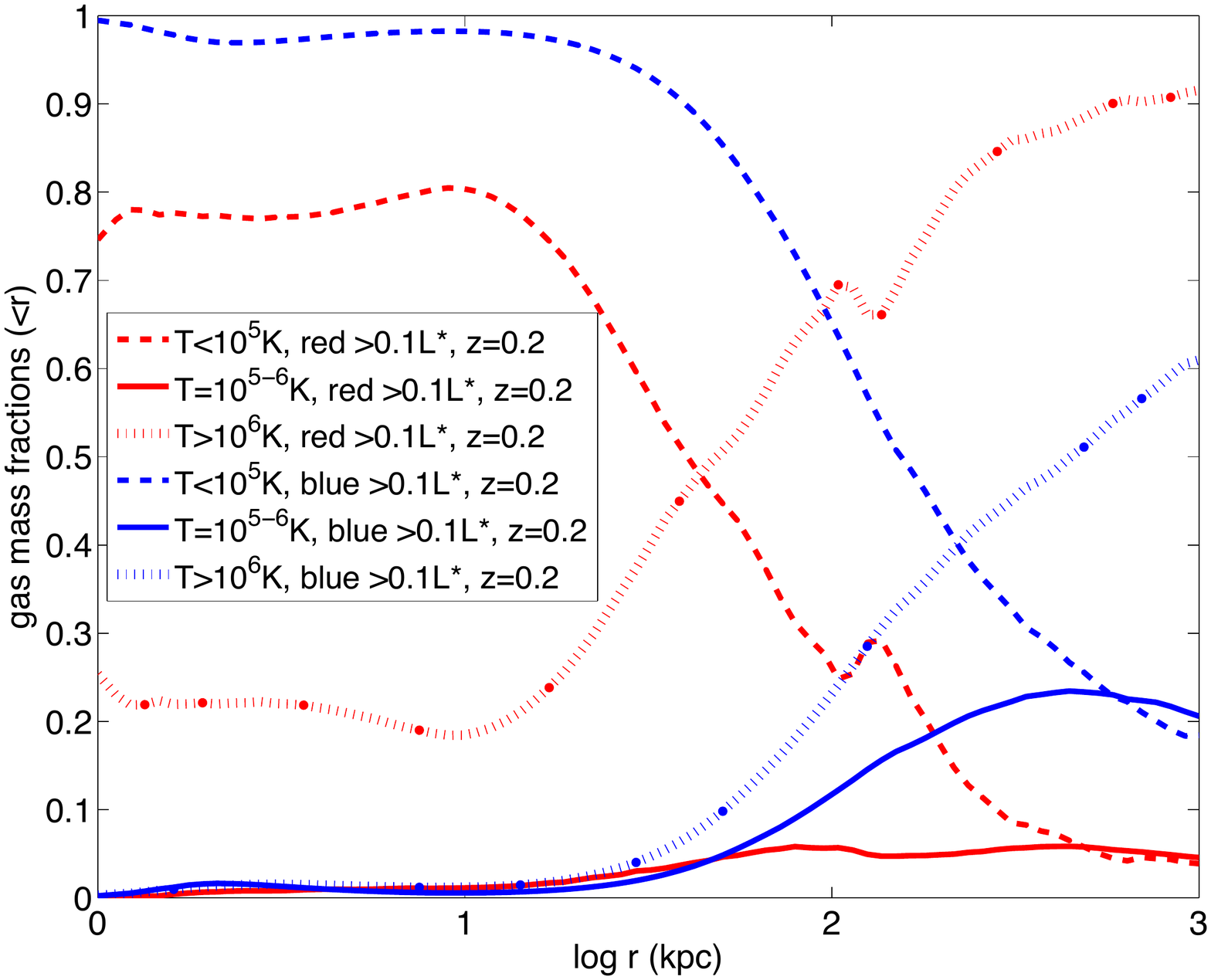}}     
\end{center}
\vskip -0.5in
\caption{
shows the differential (left panel) and cumulative (right panel) gas mass fractions as a function of radius 
for cold (dashed curves), warm-hot (solid curves)
and hot gas (dotted curves) around {\color{blue} blue (blue curves)}
and {\color{red} red (red curves)} $>0.1_L*$ galaxies at $z=0.2$.
}
\label{fig:Lr011e4gascompperc}
\end{figure}

\begin{figure}[h!]
\begin{center}
\vskip -0.0cm
\centering
\hskip -0.45in
\resizebox{3.7in}{!}{\includegraphics[angle=0]{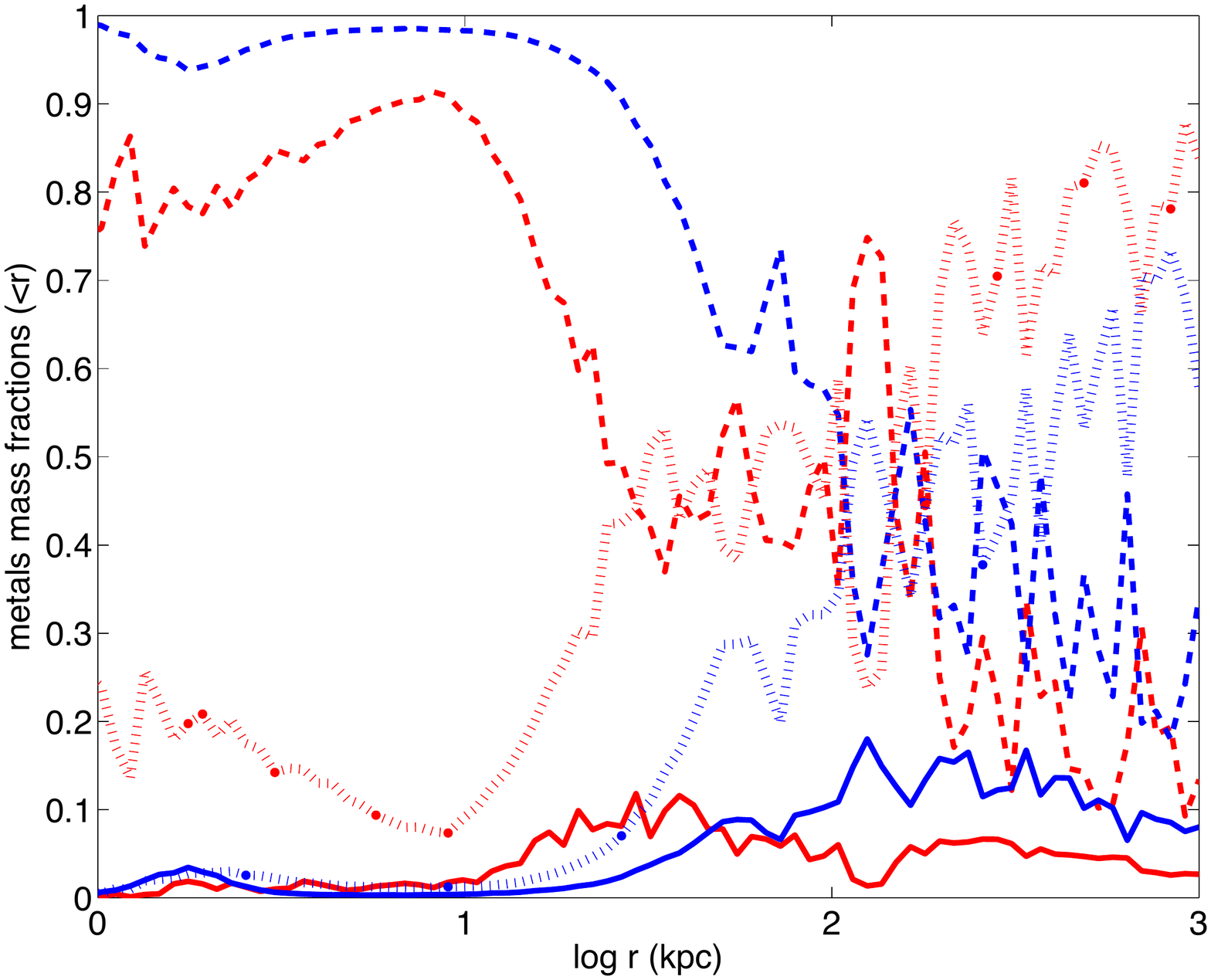}}    
\hskip -0.5in
\resizebox{3.7in}{!}{\includegraphics[angle=0]{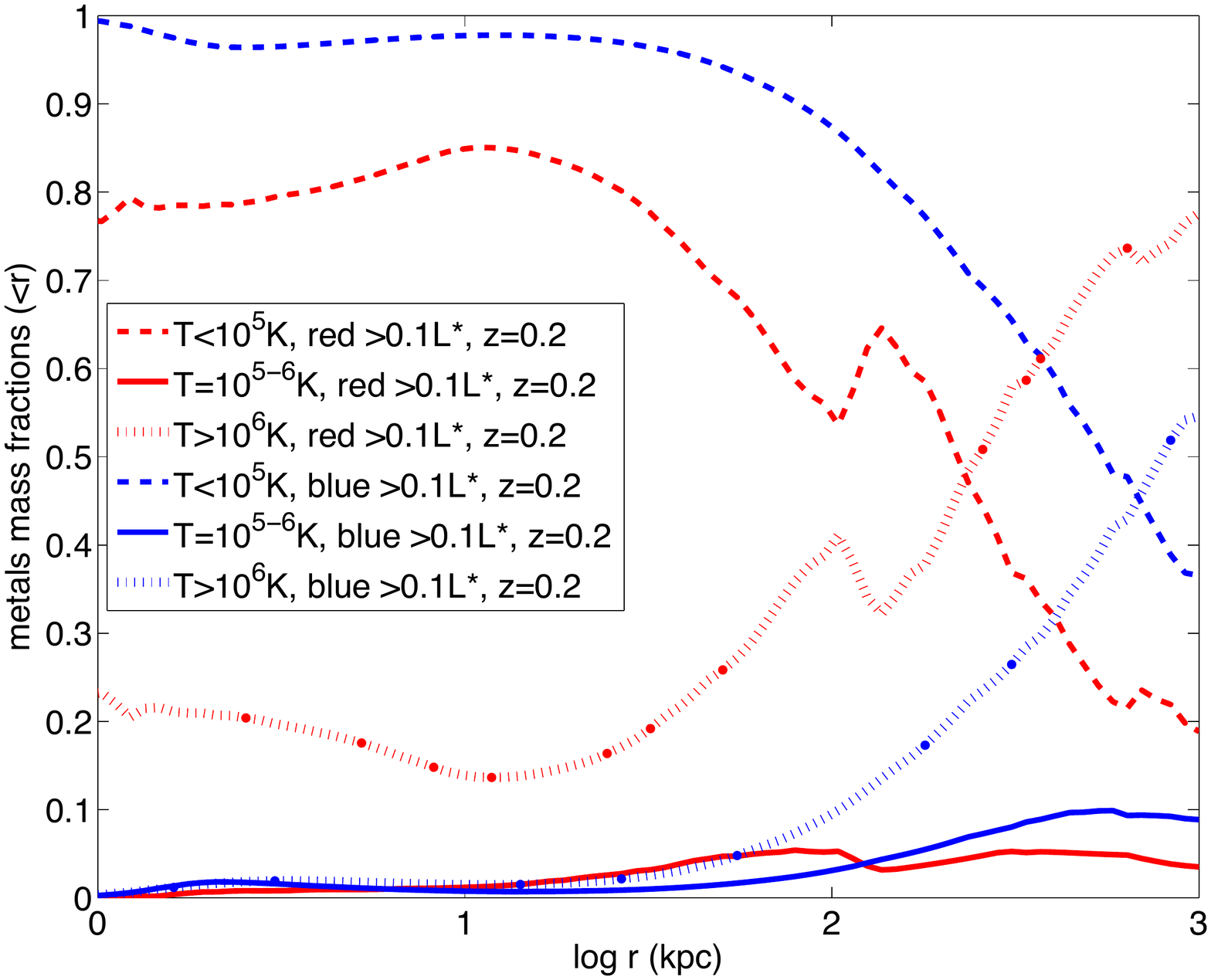}}        
\end{center}
\vskip -0.5in
\caption{
shows the differential (left panel) and cumulative (right panel) gas metals mass fractions as a function of radius 
for cold (dashed curves), warm-hot (solid curves)
and hot gas (dotted curves) around {\color{blue} blue (blue curves)}
and {\color{red} red (red curves)} galaxies at $z=0.2$.
}
\label{fig:Lr011e4mtlcompperc}
\end{figure}

Figure~\ref{fig:Lr011e4gascompperc} shows the differential (left panel) 
and cumulative (right panel) mass fractions of each gas component as a function of 
galactocentric distance for {\color{red} red (red curves)} and {\color{blue} blue (blue curves)} galaxies.
We note that the fluctuating behaviors (mostly in the differential functions on the left panel)
are due to occasional dense cold clumps in neighboring galaxies.
Overall, we see that within about (10,30)kpc for ({\color{red} red}, {\color{blue} blue}) galaxies the cold ($T<10^5$K) gas component
completely dominates, making up about (80\%, $>95$\%) of all gas at these radii.
For both $>0.1L_*$ ({\color{red} red}, {\color{blue} blue}) galaxies 
cold gas remains the major component up to $r=({\color{red}30},{\color{blue}150})$kpc,
within which its mass comprises 50\% of all gas.
At $r>({\color{red}30},{\color{blue}200})$kpc for ({\color{red}red}, {\color{blue}blue}) galaxies
the hot ($T>10^6$K) gas component dominates.
The warm gas component, while having been extensively probed observationally,
appears to be a minority in both {\color{red} red} and {\color{blue} blue} galaxies at all radii.
The warm component's contribution to the overall gas content reaches its peak value
of {\color{blue}$\sim 30\%$} at $r=100-300$kpc for {\color{blue} blue} galaxies,
whereas in {\color{red} red} galaxies it is negligible at $r<10$kpc and hovers around {\color{red}5\%} level at $r=30-1000$kpc.
The prevalence of cold gas at small radii in {\color{red} red} (i.e., low star formation activities) galaxies
is intriguing and perhaps surprising to some extent.
Some recent observations indicate that early-type galaxies in the real universe
do appear to contain a substantial amount of cold gas, consistent with our findings.
For example, \citet[][]{2012Thom} infer a mean mass of $10^9-10^{11}\msun$ of gas with $T<10^5$K at $r<150$kpc
based on a sample of $15$ early-type galaxies at low redshift from COS observations,
which is shown as the red square (its horizontal position is slightly shifted to the right for
display clarity) in Figure~\ref{fig:Lr011e4gascompgasmass}.
Their inferred range is in fact consistent with our computed value of $\sim 6\times 10^{10}\msun$ 
of cold $T<10^5$K gas for {\color{red} red} $>0.1L_*$ galaxies shown as the {\color{red} red dashed curve} in 
Figure~\ref{fig:Lr011e4gascompgasmass}.

\begin{figure}[h!]
\begin{center}
\vskip -0.0cm
\centering
\hskip -0.2in
\resizebox{4.5in}{!}{\includegraphics[angle=0]{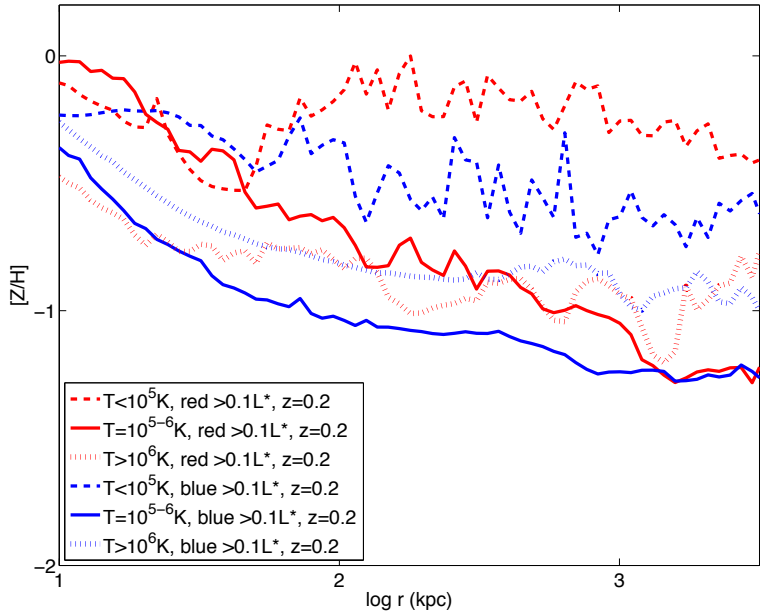}}    
\end{center}
\vskip -0.5in
\caption{
shows as a function of radius the metallicity 
for cold (dashed curves), warm-hot (solid curves)
and hot gas (dotted curves) around {\color{blue} blue (blue curves)}
and {\color{red} red (red curves)} galaxies at $z=0.2$.
}
\label{fig:ZmtlprofL011e4}
\end{figure}

Figure~\ref{fig:Lr011e4mtlcompperc} that is analogous to Figure~\ref{fig:Lr011e4gascompperc} 
shows the corresponding distributions for metals mass fractions in the three components
for {\color{red} red} and {\color{blue} blue} galaxies.
The overall trends are similar to those for total warm gas mass.
We note one significant difference here.
The overall dominance of metals mass in cold gas extends further out radially for both 
{\color{red} red} and {\color{blue} blue} galaxies, whereas the contributions to metal mass from the other two components 
are compensatorily reduced.
For example, we find that the radius within which the cold mass component makes up 50\% of total gas 
in ({\color{red} red}, {\color{blue} blue}) galaxies is (40,150)kpc,
where the radius within which the cold mass component makes up 50\% of total gas metals  
in ({\color{red} red}, {\color{blue} blue}) galaxies becomes (200,500)kpc.
This is largely due to a significantly higher metallicity of the cold component in both {\color{red} red} and {\color{blue} blue} galaxies
compared to the other two components, as shown in Figure~\ref{fig:ZmtlprofL011e4}.
We also note from Figure~\ref{fig:ZmtlprofL011e4} that
the warm gas in {\color{red} red} galaxies has a higher metallicity than in {\color{blue} blue} galaxies,
as found earlier, with the mean metallicity 
($\sim 0.25\zsun$, $\sim 0.11\zsun$) in ({\color{red} red}, {\color{blue} blue}) galaxies within a radius of $150$kpc.

\section{Conclusions}

The distribution and evolution of the intergalactic medium 
have largely been addressed by cosmological simulations
\citep[e.g.,][]{1999Cen, 2001Dave, 2011Cen, 2012bCen}.
The global distribution and composition of halo gas in and around galaxies at low redshift ($z<0.5$)
are addressed here, utilizing
state-of-the-art high-resolution ($460h^{-1}$pc), large-scale 
cosmological hydrodynamic simulations with validated star formation and feedback prescriptions.

We find that within about ({\color{red}10},{\color{blue}30})kpc for ({\color{red} red}, {\color{blue} blue}) $>0.1L_*$ galaxies the cold ($T<10^5$K) gas component
is the primary gas component making up about (80\%, $>95$\%) of all gas.
For both $>0.1L_*$ ({\color{red} red}, {\color{blue} blue}) galaxies 
cold gas remains the major component up to $r=({\color{red}30},{\color{blue}150})$kpc,
within which its mass comprises 50\% of all gas.
At $r>({\color{red}30},{\color{blue}200})$kpc for ({\color{red}red}, {\color{blue}blue}) galaxies
the hot ($T>10^6$K) gas component dominates.
The warm ($T=10^5-10^6$K) gas component makes a minor contribution 
to the overall gas mass as well as gas metal mass in both {\color{red} red} and {\color{blue} blue} galaxies.
The warm component's contribution to the overall gas content reaches its peak value
of {\color{blue}$\sim 30\%$} at $r=100-300$kpc for {\color{blue} blue} galaxies,
whereas in {\color{red} red} galaxies its contribution is negligible at $r<10$kpc and capped at {\color{red}5\%} level at $r=30-1000$kpc.
Where comparisons with observations are possible, we find that 
the amount of warm gas and and oxygen mass in star forming galaxies 
are in agreement with observations \citep[e.g.,][]{2011Tumlinson},
so is the amount of cold gas in early-type galaxies at low redshift \citep[e.g.,][]{2012Thom}.
The presence of a significant amount cold gas in {\color{red} red} galaxies at low redshift is new and somewhat surprising.
The nature of this cold gas and its role in star formation in {\color{red} red} galaxies
are not addressed in the present paper.
This and signatures of the predicted dominance of hot gas in blue galaxies (as well as red galaxies) at large radii will be addressed elsewhere.

In addition to the agreement between our simulations and observations
with respect to the global O~VI incidence rate \citep[][]{2012bCen},
we show that our predicted correlations between galaxies and strong O~VI absorbers of 
column density $N_{\rm OVI} \ge 14~$cm$^{-2}$ are in excellent agreement with observations
with $\chi$ square per degree of freedom equal to $1.2$.
On the other hand, when we use only photoionized O~VI absorbers ($T<3\times 10^4$K) in the simulations,
the comparisons between simulations and observations become substantially less favorable
with $\chi$ square per degree of freedom equal to $7.6$;
the significant disagreement stems from the photoionized O~VI absorbers being too distant from galaxies.
The O~VI line incidence rate per $\ge 0.1L_*$ galaxy around {\color{blue} blue} ($g-r < 0.6$) galaxies in our simulations
is higher than that around $\ge 0.1 L_*$ {\color{red} red} ($g-r>0.6$) galaxies by a factor of $\sim 4$ at $r\le 100-300$kpc, 
increasing to $\ge 10$ at $r\le 20$kpc, in reasonable agreement with extant observations \citep[e.g.,][]{2009Chen,2011Prochaska},
whereas in the photoionization dominated model the ratio is zero at $r<100~$kpc.
Thus, the cross correlations between galaxies and O~VI lines provide powerful differentiation between collisional and photo ionization models,
with collisional ionization dominance for strong ($N_{\rm OVI}\ge 10^{14}$cm$^{-2}$) O~VI absorbers being favored currently.

The O~VI-bearing halo gas (i.e., the warm component) is found to be {\it ``transient"} in nature and hence requires constant sources.
We perform analysis to unravel the sources and their relations to galaxy formation.
We find that, on average, to within a factor of two,
contributions to warm metals in the halo gas from
star formation feedback ($F_{\color{red} r}$), accretion of intergalactic medium ($A_{\color{red} r}$)
and gravitational shock heating ($G_{\color{red} r}$) are $(F_{\color{red} r},A_{\color{red} r},G_{\color{red} r})=({\color{red}30\%, 30\%, 40\%})$ for {\color{red} red} $\ge 0.1L_*$ galaxies at $z=0.2$.
For {\color{blue} blue} $\ge 0.1L_*$ galaxies at $z=0.2$ contributions 
are $(F_{\color{blue} b},A_{\color{blue} b},G_{\color{blue} b})=({\color{blue}48\%, 48\%, 4\%})$.
For both red and blue galaxies, the amounts of warm gas in inflows and outflows are comparable.
The mean metallicity of warm halo gas in ({\color{red} red}, {\color{blue} blue}) galaxies 
is ($\sim 0.25\zsun$, $\sim 0.11\zsun$).
Environmental dependence of O~VI-bearing halo gas is as follows.
In low density environments the metallicity of inflowing warm gas 
is substantially lower than that of outflowing warm gas;
the opposite is true in high density environments.

\smallskip
\acknowledgements{
I would like to thank Dr. M.K.R. Joung for help on
generating initial conditions for the simulations and running a portion
of the simulations and Greg Bryan for help with Enzo code,
Drs. John Wise and Matthew Turk for very useful help with analysis program yt\citep[][]{2011Turk},
and an anonymous referee for constructive reports.
Computing resources were in part provided by the NASA High-
End Computing (HEC) Program through the NASA Advanced
Supercomputing (NAS) Division at Ames Research Center.
The research is supported in part by NASA grant NNX11AI23G.}


\end{document}